\documentclass[aps,pre,showpacs,twocolumn]{revtex4}
\usepackage{graphicx}
\usepackage{amsmath}

\begin{document}

\title{Memory and aging effects in interacting sub-10nm nanomagnets\\ with
large uniaxial anisotropy}

\author{Kai-Cheng Zhang and Bang-Gui Liu}
\affiliation{Institute of Physics, Chinese Academy of Sciences,
Beijing 100190, China} \affiliation{Beijing National Laboratory
for Condensed Matter Physics, Beijing 100190, China}

\date{\today}

\begin{abstract}
Using a nonequilibrium Monte Carlo method suitable to
nanomagnetism, we investigate representative systems of
interacting sub-10nm grained nanomagnets with large uniaxial
anisotropy. Various magnetization memory and aging effects are
found in such systems. We explain these dynamical effects using
the distributed relaxation times of the interacting nanomagnets
due to their large anisotropy energies.
\end{abstract}


\maketitle

\section{Introduction}

Nanomagnets attract huge interest because of their amazing
properties and promising
applications\cite{nm3,nm4,SMM1,SMM-QC,SMM-st,SMM-book}. For
well-separated nanomagnets (including single-molecule magnets),
quantum tunnelling, interference, and coherence can be observed at
extra-low temperatures\cite{SMM-book,SMM-t,SMM-i,SMM-c,SMM-c1},
and the magnetization behaviours at some higher temperatures can
be described by Neel-Brown law \cite{NB,NB1}. When inter-particle
distances become small enough, the dipolar-dipolar interaction
will modify the magnetization behavior leading to some
super-spin-glasses behaviors\cite{ssg1,ssg4}, similar to
conventional spin glasses \cite{sg1,sg2,sg3,sg4}. Grained
nanomagnets with large {\it uniaxial} anisotropy are essential to
modern magnetic data storage. For typical CoCrPtB media, usual
average grain sizes must be about 10nm to keep magnetic stability
of 10 years at room temperature\cite{recording}. In the case of
CoCrPt-oxide media for perpendicular recording, dominant
inter-grain interactions are weak antiferromagnetic (AFM)
couplings and average grain size can be 8nm or smaller for the
same stability\cite{recording}. Such average size can even be
reduced down to 3nm or smaller when FePt in the L1$_0$ phase is
used as data storage media, because its magnetocrystalline
anisotropy reaches to 44 meV/nm$^3$\cite{recording}. Such
nanomagnets with large uniaxial anisotropy, especially when
composing special systems, can yield various dynamical phenomena
waiting for exploration.

Here we explore dynamical magnetic properties of representative
systems  of sub-10nm grained nanomagnets with large uniaxial
anisotropy. We use the giant spin approach for the nanomagnet
because the magnetic interactions between electronic spins in it
are strong. We assume that the magnetic anisotropy energies
satisfy a Gaussian distribution to consider their fluctuations due
to different shapes, sizes, and interfacial environments, and the
inter-nanomagnet interactions, including magnetic dipolar
interaction, are AFM\cite{recording}. We use a giant-spin model
and a dynamical spin Monte Carlo (DSMC) method\cite{dsmc} to
correctly simulate dynamical magnetization of the giant-spins of
the component nanomagnets. Through systematical DSMC simulations,
we find various field-cooling (FC) and zero-field-cooling (ZFC)
magnetization memory and aging effects in such systems. We explain
these dynamical effects uniformly in terms of the
continuously-distributed relaxation times of the component
nanomagnets due to the various uniaxial anisotropy energies. More
detailed results will be presented in the following.

\section{Model, method and parameters}

We consider a finite two-dimensional lattice of grained
nanomagnets. Each nanomagnet actually includes many magnetic
atoms, but the magnetic interactions between the magnetic atoms
are much stronger than those among different nanomagnets. The
inter-nanomagnet magnetic interactions are described by an
antiferromagnetic coupling according to actual materials for
modern magnetic data storage\cite{recording}. We take the giant
spin approach and use one spin variable $\vec{S}_i$ to describe
the magnetic property of each nanomagnet. Because the spin value
$S_i$ of such a nanomagnet is typically $10^2\!\sim\! 10^3$, large
anisotropy energies of such nanomagnets are mainly dependent on
their shapes and interfacial environments\cite{recording}, and
thus can be reasonably assumed to satisfy a Gaussian distribution.
This simplification keeps the main physics of these nanomagnet
systems. Because $S$ is large enough, we treat the spin operator
$\vec{S}_i$ as $S_i\vec{s}_i$, where $\vec{s}_i$ is a classical
unit vector. Generally speaking, $S_i$ should vary from one
nanomagnet to another, but for our systems all the $\{S_i\}$ are
uniform enough to be let have the same average value $\bar{S}$
because the deviations from $\bar{S}$ make little differences.
Therefore, our model can be described by the Hamiltonian,
\begin{equation}
H=-\sum_i k_{ui} \vec{s}^{\small{~}2}_{iz}+\sum_{i,j}
J_{ij}\vec{s}_i\cdot \vec{s}_j-\gamma \vec{B} \cdot \sum_i
\vec{s}_i
\end{equation}
where $k_{ui}$ describes the uniaxial anisotropy energy satisfying
a Gaussian distribution with the average value $k_u$ and width
$\sigma_u$, $\gamma$ is defined as $g\mu_0 \mu_B\bar{S}$, and the
field $\vec{B}$ is in the easy axis. Here $J_{ij}$ describes the
inter-nanomagnet AFM interactions. For actual perpendicular media
for modern magnetic data storage, we assume that the easy axis is
perpendicular to the plane of the $N\times N$ nanomagnet lattice.
In such a setup, the magnetic dipolar interaction is reduced to an
AFM inter-nanomagnet interaction, which has been included in the
$J_{ij}$ parameters in Eq. (1).

Each of the spins has two meta-stable orientations along the easy
axis and needs to overcome an energy barrier to achieve a
reversal. We use the DSMC method to simulate the spin dynamics of
these nanomagnet systems\cite{dsmc}. This method originates from
the kinetic Monte Carlo method for simulating atomic dynamics
during epitaxial growth\cite{kmc}. Using $\theta_i$ to describe
the angle deviation of $\vec{s}_i$ from the easy axis, we express
the energy increment of the $i$-th nanomagnet as $\Delta E_i=
k_{ui} \sin^2\theta_i -h_i(\cos\theta_i-1) $ to leading order,
where $h_i=(\gamma B-\sum_jJ_{ij}s_j)s_i$ and the reduced variable
$s_i$ takes either 1 or -1\cite{dsmc}. As a result, an energy
barrier $\Delta E_i=(2k_{ui}+h_i)^2/4k_{ui}$ must be overcome to
achieve the reversal of the $i$-th spin. The rate for the spin
reversal obeys Arrhenius law $R_i=R_0\exp(-\Delta E_i/k_B T)$,
where $k_B$ is Boltzmann constant and $T$ is temperature.

In our simulation, we use a typical value $1.0\times 10^9$/s for
the characteristic frequency $R_0$. As for the anisotropy energy
parameters, we reasonably assume that the average value $k_u$ of
$\{k_{ui}\}$ is 80.0 meV and the Gaussian width $\sigma_u$ is 44.7
meV. We change temperature $T$ by a step 0.2 K and set the
sweeping rate to 1 K/$s$ in most of the following cases. Special
cases will be explicitly described otherwise. We take $N=40$ and
use periodic boundary condition. This lattice size is appropriate
considering the actual situation in the media for the modern
magnetic data storage. It almost does not matter which boundary
condition is chosen with the lattice size. Further simulations
with larger $N$ are done for confirmation. Each of our data is
obtained by calculating the average value over 500$\sim$1000
independent simulation runs.

\section{Main simulated results}

{\it FC and ZFC magnetization curves}. We simulate the FC curves
by calculating the average magnetizations under 100 Oe at every
$T$ point when cooling from 110 to 10 K. The ZFC curves are
simulated by letting the system cool under zero field from 110 to
10 K and then calculating the average magnetizations under 100 Oe
at every $T$ point when warming from 10 to 110 K. Our simulation
realizes what happens in measuring experimental FC and ZFC
magnetization curves. Our simulated results are presented in
Figure 1. For zero interaction or $J=0$, the FC magnetization
increases monotonously with decreasing $T$, and the ZFC
magnetization increases first with increasing $T$, reaches its
maximum at the blocking temperature $T_B=55$ K, and then
decreases. The FC and ZFC magnetization curves follow the Curie
law above the temperature $T_m=70$ K. The difference between the
FC and ZFC magnetizations diminishes above $T_m$. Such behaviors
are key features of nanomagnets, showing up in super-spin glass
systems\cite{ssg1} and interacting nanoparticles\cite{m1,m3}. When
$J$ is larger, the magnetization in the FC and ZFC curves becomes
substantially smaller, as shown in Figure 1. When $J$ becomes
further larger, a minimal magnetization can be seen below $T_B$.
It is at 30 K for 2.0 meV. When $J$ is less than 0, the
magnetization in such curves becomes larger but the curve shape
remains nearly the same.

\begin{figure}[!h]
\begin{center}
\includegraphics[width=8.5cm]{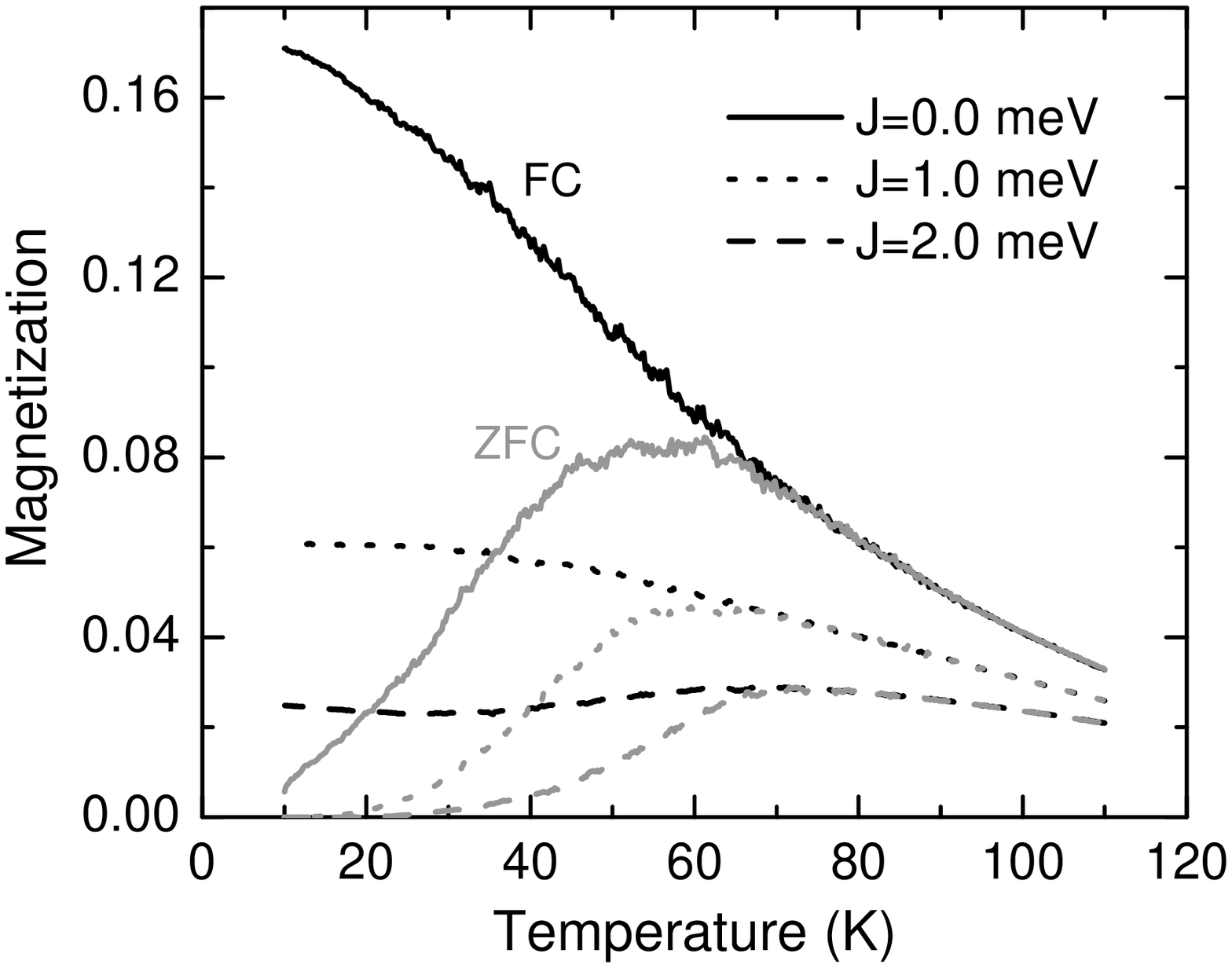}
\caption{Simulated FC (black) and ZFC (gray) magnetization curves
under a field 100 Oe for $J=$ 0.0, 1.0, and 2.0 meV. These curves
are calculated in the same way as corresponding experimental
magnetization curves are measured.}
\end{center}
\end{figure}

\begin{figure}[!h]
\begin{center}
\includegraphics[width=8.5cm]{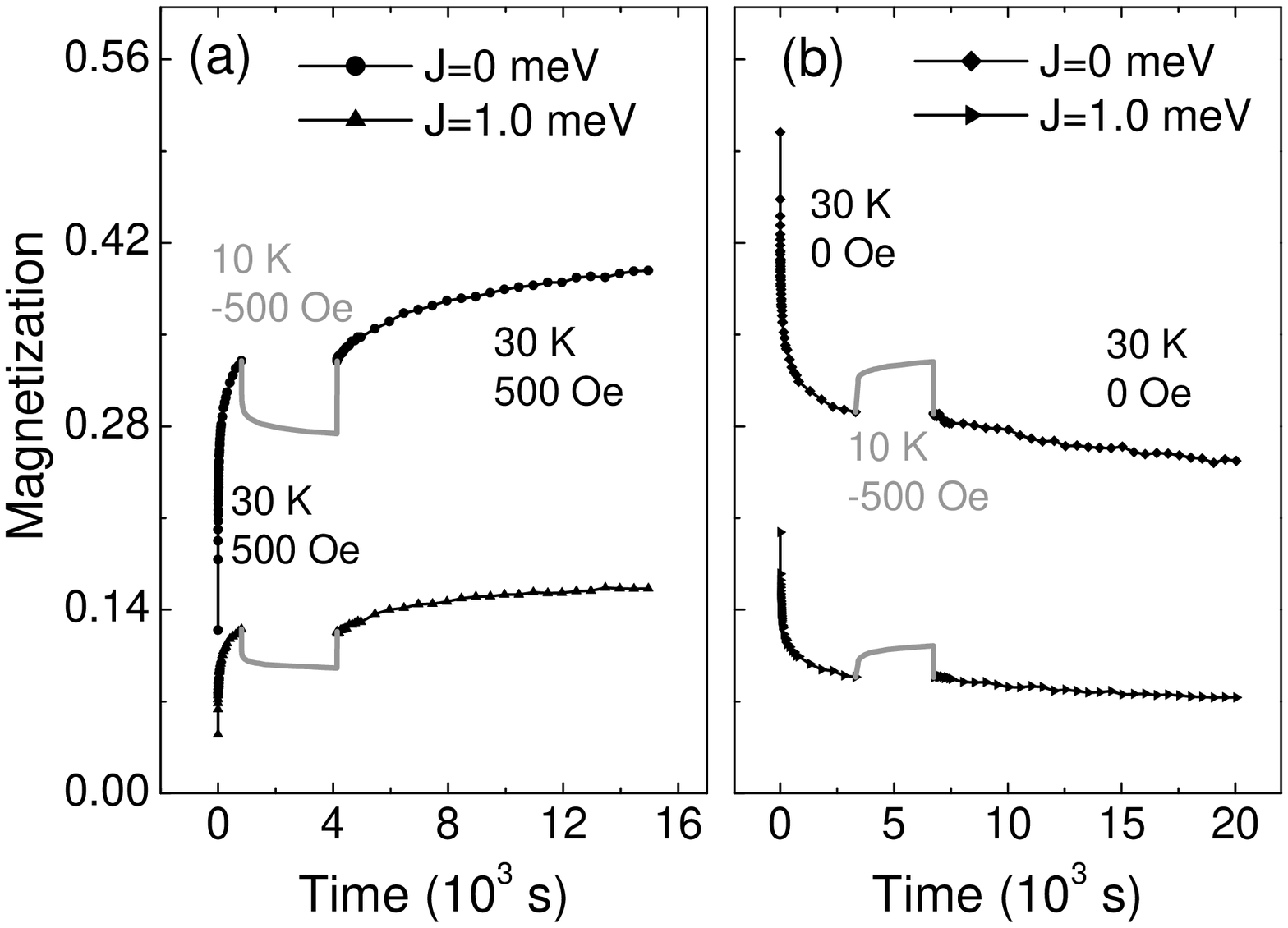}
\caption{Simulated ZFC (a) and FC (b) memory effects over time
with $J=0$ and $1.0$ meV. (a) The temperature is 30 K and the
field 500 Oe at all the time points except between $t_1=814$ s and
$t_2=4129$ s (gray); in the special time period, the temperature
and field are changed to 10 K and -500 Oe. (b) The temperature is
30 K and the field 0 Oe at all the time points except between
$t'_1=3416$ s and $t'_2=6730$ s (gray); in the special time
period, the temperature and field are changed to 10 K and -500 Oe.
}
\end{center}
\end{figure}

{\it Simulated ZFC and FC memory effects over time}. Cool the
system under zero field to a low temperature, 30 K, and then apply
a field 500 Oe and let the system relax from $t=0$. At $t_1=814$
s, we change $T$ to 10 K and reverse the field, and at
$t_2=4124$s, we recover the original temperature and field. We
calculate the average magnetization when the system relaxes with
time from $t=0$ to $t=15000$s, and present the results in Figure
2a. The magnetization at $t_2$ is equivalent to that at $t_1$,
that is, the system keeps the memory at $t_1$ although it
undergoes relaxation from $t_1$ to $t_2$ under the different
temperature and field. This is a ZFC memory effect over time. In
addition, we cool the system under a field 500 Oe to 30 K, and
then remove the field and let the system relax from $t=0$, but we
change the temperature to 10 K and the field to -500 Oe between
$t'_1=3416$s and $t'_2=6730$s. Meanwhile we calculate the
magnetization from $t=0$ to $t=20000$s. The result, shown in
Figure 2b, shows an FC memory effect over time,
$m_{t'_2}=m_{t'_1}$. Similar effect was observed in super-spin
glass systems and interacting nanoparticles\cite{ssg4,m1,m3}. Our
further calculations show that a small inter-spin interaction,
either AFM or FM, does not affect the effects substantially.

\begin{figure}[!h]
\begin{center}
\includegraphics[width=8.5cm]{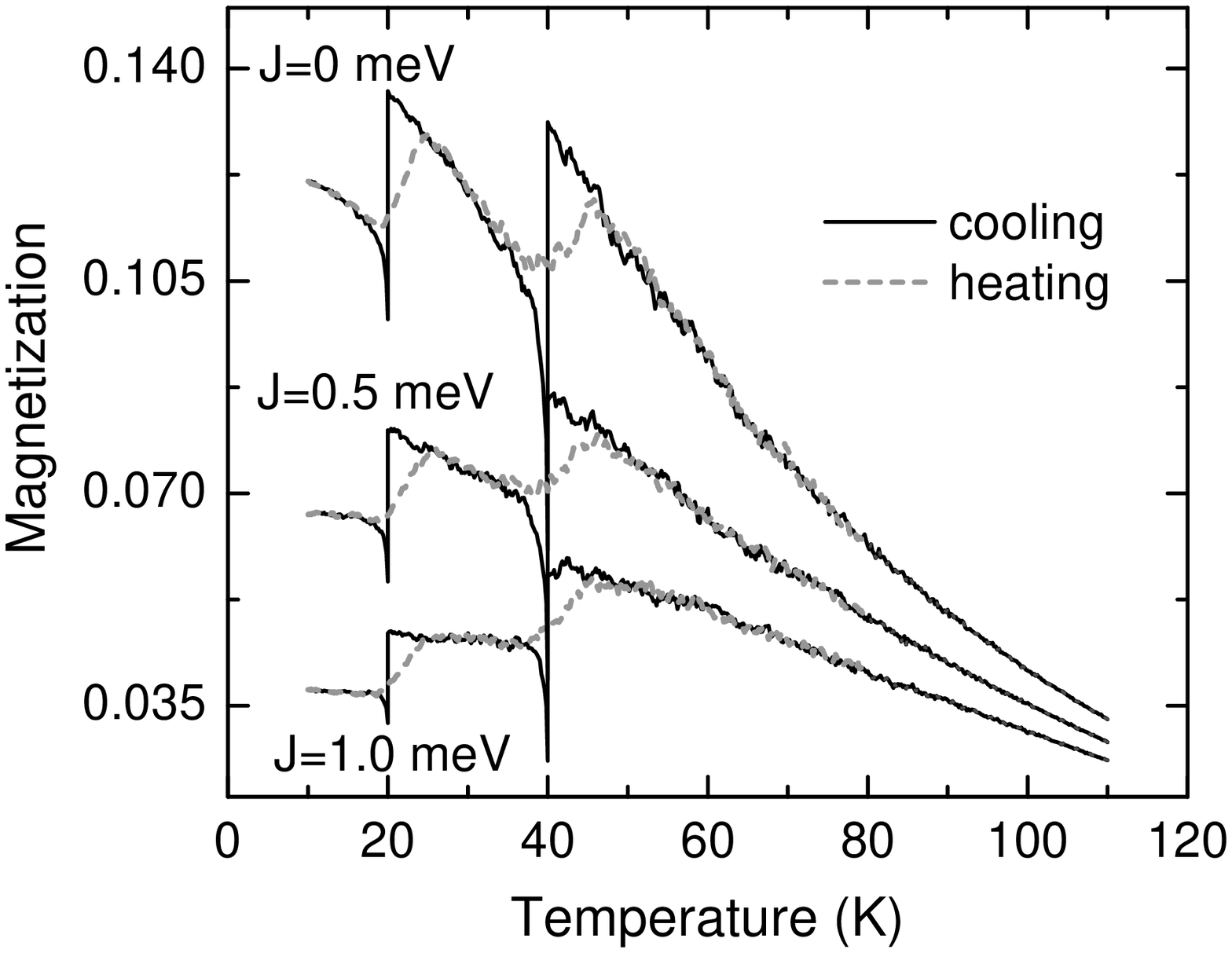}
\caption{Simulated FC memory effects over temperature for $J=$
0.0, 1.0, and 2.0 meV. During cooling (black), the temperature is
changed homogeneously and the field is 100 Oe at all the
temperature points except 40 K and 20 K. At these two special
temperature points, the system is kept at the temperature for
additional 40 s and 60 s, respectively, and the field is kept
constantly the same -100 Oe meanwhile. The heating (gray dash) is
homogeneous in time under the same field.}
\end{center}
\end{figure}

\begin{figure}[!h]
\begin{center}
\includegraphics[width=8.5cm]{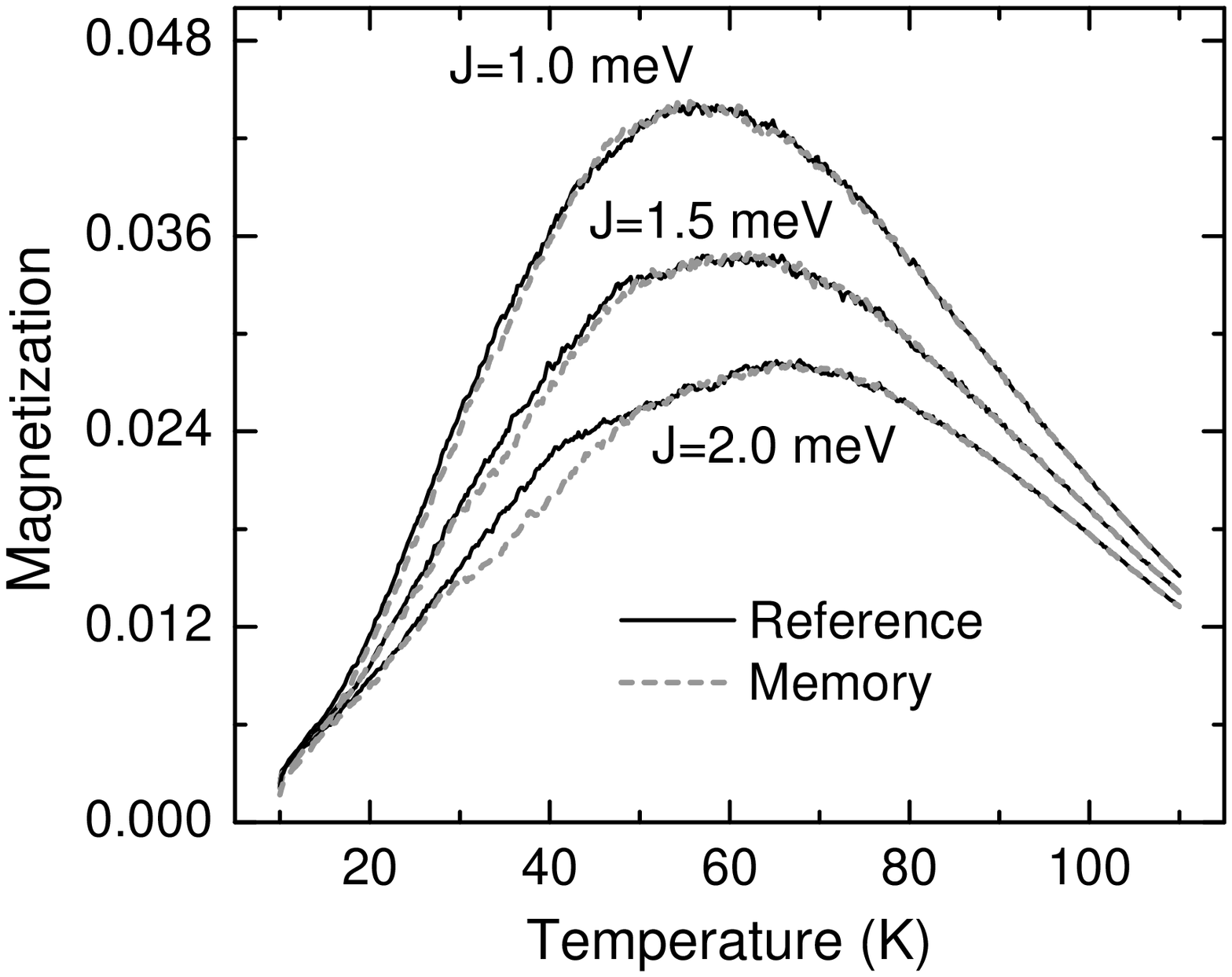}
\caption{Simulated ZFC memory effects over temperature for $J=$
1.0, 1.5, and 2.0 meV. The references (black lines) are standard
ZFC magnetization curves with 100 Oe, and the memory curves (gray
dash lines) are made by keeping the system relaxing for additional
100 s at 40 K during the cooling. The heating is always
homogeneous in time and field.}
\end{center}
\end{figure}

{\it Simulated FC memory effect over temperature}. Cool the system
under a field 100 Oe from 110 to 10 K, but at two temperature
points, 40 K and 20 K, the field is reversed and then recovers
after 40 s and 60 s, respectively. The resulting FC magnetization
curves for $J=0$, $0.5$, and $1$ meV are shown in Figure 3. The
step is clearly seen at both 40 K and 20 K for each of the three
curves. The abrupt magnetization change is because the field is
kept as the reversed value, -100 Oe, for additional 40 s or 60 s.
When the cooling ends at 10 K, we warm the system under 100 Oe
from 10 to 110 K. The corresponding three warming magnetization
curves are presented in Figure 3. Clear magnetization drop is
observed at both 40 K and 20 K, although the warming is
homogeneous in time and field. This means that the system during
the warming has the memory of the abrupt magnetization changes at
the same temperatures during the cooling. Similar effects have
been observed in interacting nanoscale systems\cite{ssg4,m1,m3}.
Our results show that a small inter-spin interaction changes the
FC memory effect only a little.

{\it Simulated ZFC memory effect over temperature}. The above
effects are essentially independent of inter-spin interactions
$J$. Such interactions weaken, even can break these memory
effects, but appropriate $J$ values can lead to another memory
effect shown in Figure 4. First, cool the system homogeneously
under zero field, calculate standard ZFC magnetization curves when
warming the system under 100 Oe, and take them as references;
then, cool the system under zero field but keep it for additional
100s when $T=40$ K, calculate corresponding magnetization curves
when warming the system under 100 Oe. Although the warming is
homogeneous in time, the magnetization near $40$ K is smaller,
which implies that the system has the memory of the waiting at 40
K during the cooling. These are consistent with observations in
the cases of other interacting super-spin systems\cite{exp1,exp2}.
It is clear that nonzero $J$ values are helpful for the memory
effect.

\begin{figure}[!h]
\begin{center}
\includegraphics[width=8.5cm]{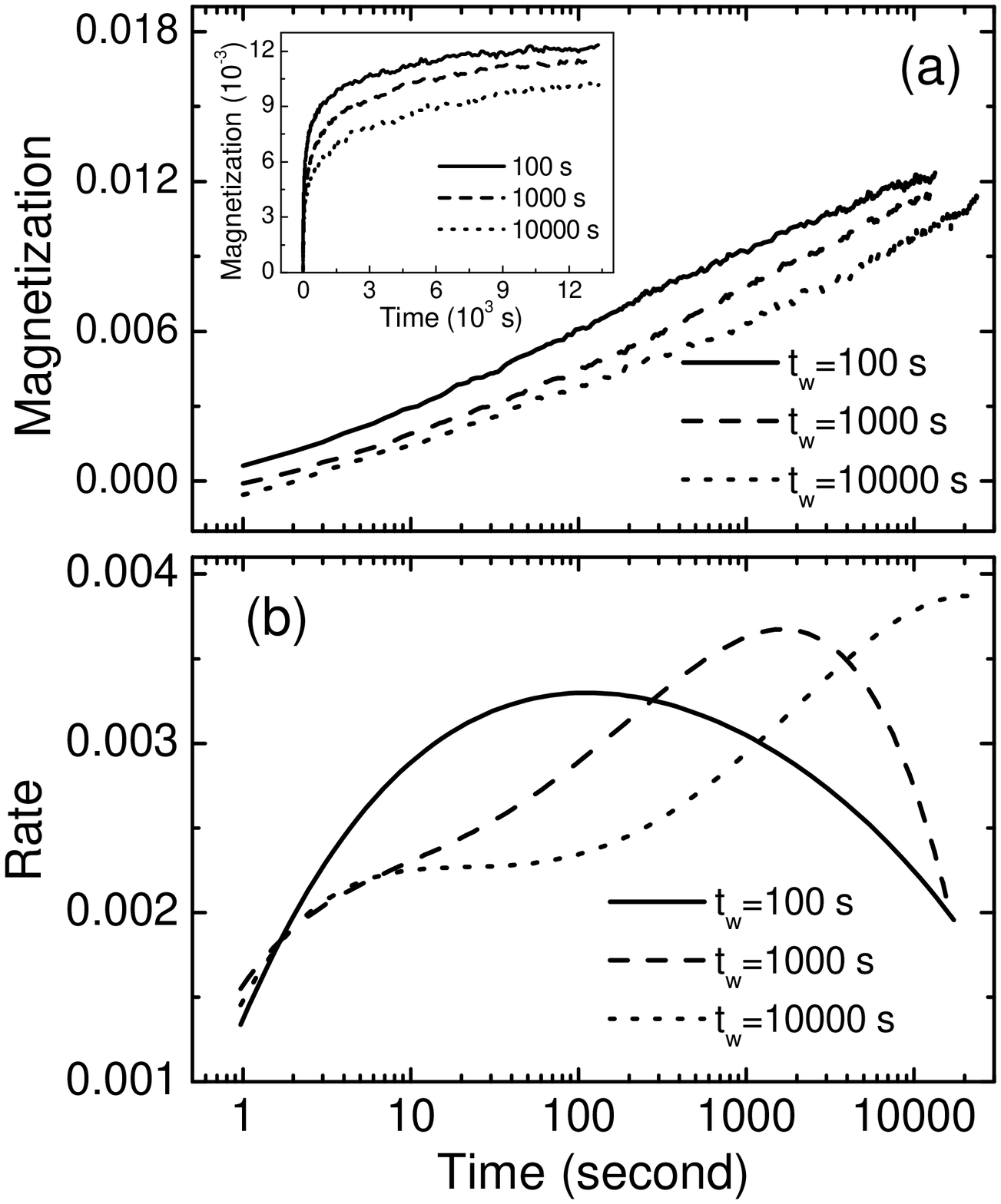}
\caption{Simulated aging effect. The magnetization $m$ (a) and
rate $S$ (b) as functions of time $t$ at 15 K under 100 Oe for
three waiting time periods $t_w$: 100s (solid), 1000s (dash), and
10000s (dot). The inset in (a) shows the same $m$ results in the
linear scale of $t$.}
\end{center}
\end{figure}

{\it Simulated aging effect}. Cool the system under zero field to
15 K (below $T_B$) and keep it unchanged for a period $t_w$, and
then let it relax with time $t$ under 100 Oe. During the
relaxation, we calculate the magnetization $m$ and the rate
$S=\partial m/\partial \log t$ as functions of $t$. Our simulated
results in logarithmic scale with $t_w=100$s, $t_w=1000$s, and
$t_w=10000$s are presented in Figure 5. The inset in the upper
panel shows the magnetization in linear scale of $t$. It is clear
that the longer the waiting time, the smaller the magnetization,
and the relaxation rate has a maximum at $t\approx t_w$. This is
consistent with well-known aging effect observed in spin-glass and
interacting super-spin systems\cite{ssg1,sg1,exp1,exp2}.

\section{Dynamical mechanism and comparison with experiment}

{\it Mechanism for the memory and aging effects}. Our systems
consist of nanomagnets with large uniaxial magnetic anisotropy
which follows Gaussian distribution. The distribution of
anisotropy energy causes various relaxation times through relation
$\tau_i=\tau_0 \exp(\Delta E_i/k_BT)$. The memory and aging
effects can be explained in a unified way by the
continuously-distributed relaxation times $\{\tau_i\}$ of
nanomagnets. The key point is that for a given time period and at
a given temperature, most effectively relaxed are the nanomagnets
with appropriate relaxation time $\tau$ because those with smaller
$\tau$ have been sufficiently relaxed and those with larger $\tau$
remain frozen. For the memory effects over time shown in Figure 2,
the temperature between the two time points, $t_1$ and $t_2$ (or
$t'_1$ and $t'_2$), is substantially lower, and nothing important
happens meanwhile, although the field is changed, because the
nanomagnets with much smaller $\tau$ have already been thoroughly
relaxed, and hence the magnetization is recovered after the field
and temperature are restored. As for the memory effect shown in
Figure 3, the additional relaxation under the reversed field at
both 40 K and 20 K during the cooling makes corresponding
nanomagnets orient in the reversed direction and most of them,
remaining frozen at the lower temperatures in the further cooling,
cause the magnetization dips at both 20 K and 40 K in the warming
curve, the FC memory effect over temperature. For the ZFC memory
effect shown in Figure 4, the additional 100 s relaxation at 40 K
during the cooling, in the presence of the AFM inter-nanomagnet
coupling, makes the nanomagnets with corresponding anisotropy
energy over-relaxed so that the magnetization curve (warming from
10 K) has a small dip at the same 40 K with respect to the
reference curve. As for the aging effect shown in Figure 5, the
key point is that the system is still relaxing at the base
temperature 15 K during the additional waiting period $t_w$. For
larger $t_w$, the system becomes more relaxed and the
magnetization is a little smaller until a time $t\approx t_w$.
However, the effect of the additional relaxation of $t_w$ will
finally disappear for $t\gg t_w$. Therefore, there is a peak for
the rate of increasing the magnetization at a time $t\approx t_w$.

{\it Compared with experiments and other theories}. Similar FC and
ZFC magnetization memory effects were observed in both spin
(including super-spin) glass systems and isolated nanomagnets, and
similar ZFC memory and aging effects in spin glass
systems\cite{ssg1,ssg4,sg1,sg2,sg3,sg4,m1,m3}. Usually, memory
effects in isolated nanomagnets were attributed to broad
distribution of relaxation times for different nanomagnets. The
memory and aging effects in spin glass systems can be understood
in terms of droplet model and hierarchy
model\cite{sg1,sg2,sg3,sg4}. For our systems the magnetic
anisotropy is uniaxial and the inter-spin interaction is uniformly
AFM, and therefore our model, including inter-spin interaction, is
distinguished from other ones for those
systems\cite{ssg1,ssg4,sg1,sg2,sg3,sg4,m1,m3}.

\section{Conclusion}

Using the dynamical spin Monte Carlo method, we investigate
various nonequilibrium dynamical magnetization behaviors of the
representative systems (finite two-dimensional lattices) composed
of antiferromagnetically-coupled sub-10nm grained nanomagnets with
large uniaxial anisotropy. In each of such systems of the
nanomagnets with the same easy axis, we find both ZFC and FC
magnetization memory effects over both time and temperature and
the magnetization aging effect, which partly appeared otherwise in
different spin (or super-spin) glass systems and interacting
nanoparticles. We explain these interesting dynamical effects in a
unified way in terms of the continuously-distributed relaxation
times of the interacting nanomagnets due to their large anisotropy
energies in the actual media.

\begin{acknowledgments}
This work is supported  by Nature Science Foundation of China
(Grant Nos. 10774180, 10874232, and 60621091), by Chinese
Department of Science and Technology (Grant No. 2005CB623602), and
by the Chinese Academy of Sciences (Grant No. KJCX2.YW.W09-5).
\end{acknowledgments}

\end{document}